\documentstyle[prd,aps]{revtex}

\begin{document}

\title{Loop quantum gravity and quanta of space: 
a primer}

\author{\bf Carlo Rovelli, Peush Upadhya}
\address{Physics Department, University 
of Pittsburgh, Pittsburgh PA 15260, USA} 
\address{rovelli@pitt.edu} 
\date{\today} 
\maketitle

\begin{abstract} 
We present a straightforward and self-contained introduction to 
the basics of the loop approach to quantum gravity, and a 
derivation of what is arguably its key result, namely the 
spectral analysis of the area operator.  We also discuss the 
arguments supporting the physical prediction following this  
result: that physical geometrical quantities are quantized in a 
non-trivial, computable, fashion.  These results are not new;  
we present them here in a simple form that avoids the many 
non-essential complications of the first derivations.
\end{abstract}  

\vskip1cm

\section{Introduction}

\noindent Combining quantum mechanics and general relativity, and 
understanding the quantum properties of the spacetime geometry, 
is a central open problem in fundamental physics.  Among the 
various approaches presently pursued to address this problem (see 
\cite{Puna} for an overview), is loop quantum gravity (see 
\cite{Loops} for a recent overview).

The theoretical framework of loop quantum gravity has evolved 
through numerous twists and re-foundations 
\cite{basic,Area,Barbero,RovDP,nodes,Helvetica,altri}, and it is easy for 
the non-experts to loose track of the basics of the theory.  
Furthermore, the basic structures have been found via very 
roundabout paths, involving mathematical tools such as Penrose's 
spinor and binor calculus, Temperley-Lieb algebras, the 
Kauffman-axioms, C*-algebraic techniques, Gelfand representation 
theory, infinite dimensional measures, generalized connections, 
projective limits and so on.  Each of these tools adds a valuable 
new perspective and possible new handles to the overall picture, 
but it is now clear that none of these tools is strictly 
necessary for the definition of the basics theory and the 
derivation of the main results.  Therefore, it is now a good time 
for a simple presentation of the basics of the loop 
representation and a simple derivation of its main results.  We 
provide here such a derivation, confining ourselves to the 
definition of the (unconstrained) state space of the theory, and 
the quantization of the area.  The results we present were 
obtained in References \cite{basic,Area,Barbero,RovDP,nodes,Helvetica}.  
The derivation here is slightly original, but our aim is 
mainly to provide a simple introduction to the subject; we refer 
to the original works for some proofs and details.

We organize the presentation in two parts.  The first part 
(Section \ref{Math}) is purely mathematical: we construct an 
Hilbert space ${\cal H}$ and certain operators on ${\cal H}$, and 
we solve the spectral problem for these operators.  In the second 
part (Section \ref{Phys}), we present the motivations for a 
physical interpretation of the structures defined in the first 
part.  We add two appendices.  In the first, we recall a few 
relevant facts from $SU(2)$ representation theory, which clarify 
and make more explicit the construction presented in the main 
text; in the second, we discuss a special case disregarded in the 
main text.

\section{Mathematics} \label{Math}

In this section, we define the basic tools of the kinematics of 
the loop representation.  These will get a physical 
interpretation in Section \ref{Phys}.

\subsection{The Hilbert space}

Let $M$ be a fixed three-dimensional compact smooth manifold.  
For concreteness, we choose $M$ to be (topologically) a 
three-sphere.  Let $A$ be an $SU(2)$ connection on $M$; that is, 
$A$ is a smooth 1-form with values in $su(2)$, the Lie algebra of 
$SU(2)$.  We denote the space of the smooth $su(2)$ valued 
1-forms $A$ on $M$ as $\cal A$.  The space $\cal A$, equipped 
with the {\em sup\/} norm is a topological space.  We denote the 
space of the continuous functions $\Psi(A)$ on $\cal A$ as $L$.  
Equipped with the pointwise topology, $L$ is a topological vector 
space.

An important class of functions in $L$ is given by the 
cylindrical functions defined as follows.  Let $\Gamma$ be a 
graph embedded in $M$.  That is, $\Gamma$ is a collection of a 
finite number $n$ of (oriented) piecewise analytic curves 
embedded in $M$, which we call links and denote as 
$\gamma_{1},\ldots,\gamma_{n}$, which may overlap only at their 
end-points, which we call nodes, and denote as 
$p_{1},\ldots,p_{m}$.  The number of lines meeting in a node $p$ 
is called the valence of the node.  A node may have any valence 
equal or larger than 1.  Given a curve $\gamma$ and a connection 
$A$ defined on a manifold, the holonomy (or parallel transport 
matrix) of $A$ along $\gamma$ is defined (See Appendix A); it is 
an element of $SU(2)$, and we denote it as $U(\gamma,A)$.  Given 
a graph $\Gamma$ with links $\gamma_{1},\ldots,\gamma_{n}$, and a 
(Haar-integrable) complex function $f$ on $(SU(2))^{n}$, we 
define the function

\begin{equation} 
 \Psi_{\Gamma,f}(A)\equiv f(U(\gamma_{1},A),\ldots,
 U(\gamma_{n},A)). 
\label{cylindrical}
\end{equation}
We call functions of the form (\ref{cylindrical}) cylindrical.  
The cylindrical functions are dense in $L$.

Any cylindrical function $\Psi_{\Gamma',f'}(A)$, based on a graph 
$\Gamma'$, can be rewritten as as a cylindrical function 
$\Psi_{\Gamma,f}(A)$ based on a larger graph $\Gamma$ containing 
$\Gamma'$, by simply choosing $f$ to be independent from the links 
in $\Gamma$ but not in $\Gamma'$.  Furthermore, any link of a 
graph can be broken in two links, separated by a (bivalent) node.  
Therefore it is clear that any two given cylindrical functions 
$\Psi_{\Gamma',f'}$ and $\Psi_{\Gamma'{}',g'{}'}$ can always be 
rewritten as based on the same graph $\Gamma$ (choosing $\Gamma$ 
as the smallest graph having both $\Gamma'$ and $\Gamma'{}'$ as 
subgraphs).  Using this, we define the following quadratic form, 
defined for any two cylindrical functions.
\begin{equation}
(\Psi_{\Gamma,f},\Psi_{\Gamma,g}) \equiv 
\int_{(SU(2))^{n}} dU_{1}\ldots dU_{n}\ 
\overline{f(U_{1},\ldots,U_{n})}\ 
g(U_{1},\ldots,U_{n}). 
	\label{scalarproduct}
\end{equation}
We write this in the suggestive form 
\begin{equation}
(\Psi_{\Gamma,f},\Psi_{\Gamma,g}) = \int_{\cal A} {\cal D}A\ \,  
\overline{\Psi_{\Gamma,f}(A)}\ \Psi_{\Gamma,g}(A), 
\end{equation}
where the meaning of the infinite dimensional integration is given 
only by equation (\ref{scalarproduct}).  The quadratic form 
(\ref{scalarproduct}) defines the norm 
$||\Psi_{\Gamma,f}||^{2}\equiv (\Psi_{\Gamma,f},\Psi_{\Gamma,f})$ 
and can be extended to $L$ by continuity.  By factoring away the 
zero-norm subspace and closing $L$ in norm, we obtain a 
(non-separable) Hilbert space $\cal H$.  This Hilbert space plays 
a key role in the following.

\subsection{Gauge transformations and Diffeomorphisms}

The connection $A$ transforms under local $SU(2)$ gauge 
transformations $A \mapsto A_{V} = V^{-1}AV + V^{-1}dV$, where 
$V$ is a smooth map from $M$ to $SU(2)$.  The Hilbert space $\cal 
H$ carries a natural representation of the group of the these 
gauge transformations $\Psi(A) \mapsto \Psi(A_{V})$.  This 
representation is unitary, because the scalar product 
(\ref{scalarproduct}) is invariant.  In fact, the parallel 
trasport matrices transform as $U(\gamma,A) \mapsto 
U(\gamma,A_{V})= V(x_{i})U(\gamma,A) V(x_{f})$, where $x_{i}$ and 
$x_{f}$ are the initial and final points of $\gamma$; and the 
Haar integral in (\ref{scalarproduct}) is $SU(2)$ invariant.

The gauge invariant functions in $\cal H$ (which satisfy 
$\Psi(A)=\Psi(A_{V})$) form a proper subspace of $\cal H$, which 
we denote ${\cal H}_{0}$.

The connection $A$ transforms under a diffeomorphism $\phi: M\to 
M$ as a one form $A \mapsto \phi^*A$.  The Hilbert space $\cal 
H$ carries a natural representation of the diffeomorphism group 
$\Psi(A) \mapsto \Psi(\phi^*A)$.  This representation is unitary, 
because the scalar product (\ref{scalarproduct}) is invariant.  
In fact, the parallel trasport matrices transform as $U(\gamma,A) 
\mapsto U(\gamma,\phi^* A) = U(\phi^{-1}\cdot\gamma,A)$, where 
$[\phi^{-1} \cdot \gamma](x) = \gamma(\phi(x))$, and this 
transformation has no effect on the right hand side of 
(\ref{scalarproduct}).  Thus, the Hilbert space $\cal H$ carries 
a natural unitary representation of the group of the 
diffeomorphisms of $M$.\footnote{There are no finite norm 
diffeomorphism invariant functions in $\cal H$ (satisfying 
$\Psi(A)=\Psi(\phi^*A)$), but a space of diffeomorphism invariant 
(infinite norm) ``generalized functions'' can be naturally 
defined, and can be equipped by a natural Hilbert structure, 
using general techniques.}

\subsection{An orthonormal basis}

Consider a graph $\Gamma$ in $M$. Associate an irreducible 
representation $j_{i}$ of $SU(2)$ to each link $\gamma_{i}$ of 
the graph.  This is called coloring of the links.  We denote the 
Hilbert space on which the representation $j_{i}$ is defined as 
$H_{i}$.

Let $p$ be a $n$-valent node of $\Gamma$ and let 
$\gamma_{1}\ldots\gamma_{n}$ be the $n$ links that meet in $p$, 
and $j_{1}\ldots j_{n}$ their associated representations.  
Consider the tensor product of the Hilbert spaces of these 
representations $H_{(j_{1})}\otimes \ldots \otimes H_{(j_{n})}$.  
Let $H_{p}$ be the invariant subspace (the spin-zero component) 
of this tensor product.  Assume that $H_{p}$ has dimension equal 
or larger than one -- this amounts to restricting the valence of 
the nodes to be strictly larger than one, and to a restriction on 
the coloring of the links.  Let us fix an orthonormal basis in 
$H_{p}$.  Associate an element $v$ of this basis to the node $p$.  
Do the same for all the nodes of the graph -- that is: associate 
a basis element $v_{r}$ of $H_{p_{r}}$ to each node $p_{r}$.  
This is called coloring of the nodes.

A colored graph, namely a triplet $s=(\Gamma,j_{i},v_{r})$ is 
called an (embedded) spin network.  A spin network $s$ defines a 
cylindrical function $\Psi_{s}(A)$, called a ``spin network 
state'', as follows.  The graph of the cylindrical function is 
the graph of the spin network.  For each link $\gamma_i$, the 
quantity $j_{i}[U(\gamma_{i},A)]$ is a linear operator in the 
Hilbert space $H_{j_{i}}$.  It can be viewed as an operator from 
the $H_{j_{i}}$ associated to its initial point to the 
$H_{j_{i}}$ associated to its final point.  These operators can 
be contracted with the invariant tensors $v_{r}$ at the nodes 
(see Appendix A), obtaining a number $\Psi_{s}(A)$ that depends 
only on the spin network and the connection.

We now have the remarkable result that the spin network states 
$\Psi_{s}$ (for a fixed choice of a basis in $H_{p}$ at each 
node) form an orthogonal basis in ${\cal H}_{0}$.  This can be 
proven by direct computation, using standard $SU(2)$ 
representation theory machinery.  The normalization of the spin 
network states, to obtain an orthonormal basis, is 
simple; it is given explicitly in \cite{RovDP}. 

\subsection{The operator $E(\Sigma)$}

We now introduce some operators on $\cal H$.  It is easier to 
work with indices.  Let $x^{a}$ be local coordinates on $M$, 
$a,b,\ldots=1,2,3$ be tangent indices and $i,j,\ldots=1,2,3$ be 
indices on the Lie algebra of $SU(2)$.  Thus the components of 
the connection are given by $A(x) = A_{a}(x) 
dx^{a}= A^{i}_{a}(x) \tau^{i} dx^{a} $, 
where the $\tau^{i}$ are the three generators of $SU(2)$.  The 
first operator we define is a functional derivative ``$-i 
\frac{\delta}{\delta A_{a}^{i}(x)}$'' operator.  A functional 
derivative is a distribution and needs to be suitably smeared.  
Because of the peculiar structure of the functions in $\cal H$, 
which are (limits of sequences of) functions with support on one 
dimension, it is sufficient to smear the functional derivative in 
two dimensions only, in order to have a well defined operator.  
Let $\Sigma: \vec{\sigma}\mapsto x^{a}(\vec\sigma)$ be an 
oriented (2d) surface in $M$, and let 
$\vec\sigma=(\sigma^{1},\sigma^{2})$ be coordinates on $\Sigma$.  
We define the operator
\begin{equation}
	E^{i}(\Sigma) \equiv -i \int_{\Sigma}d\sigma^{1}d\sigma^{2} 
	\epsilon_{abc} \frac{\partial x^{a}(\vec\sigma)}{\partial \sigma^{1}} 
	\frac{\partial x^{b}(\vec\sigma)}{\partial \sigma^{2}} 
	\frac{\delta}{\delta A_{c}^{i}(x(\vec\sigma))},
	\label{E}
\end{equation}
where $\epsilon_{abc}$ is the completely antisymmetric 
tensor-density with $\epsilon_{123}=1$.  It is easy to verify 
that the operator $E^{i}(\Sigma)$ is geometrically well defined, 
that is, it is independent of the coordinates $\vec\sigma $ and 
$x^a$ chosen.

The operator $E^{i}(\Sigma)$ is well defined on the cylindrical 
functions.  To see how this may happen, consider its action on 
the spin network state $\Psi_{s}(A)$.  We denote the 
intersection points between the spin network $s$ and the 
surface $\Sigma$ as ``punctures''.  To begin with, let us 
consider the simplest case in which the surface $\Sigma$ and the 
spin network $s$ intersect on single puncture $p$, where $p$ 
lies on (the interior of) the link $\gamma$. Let $j$ be the color 
of $\gamma$.  A standard result, which can be obtained, for 
instance, by taking the first variation of the differential 
equation (\ref{holonomy}) defining the holonomy 
(see for instance \cite{Ted}), is
\begin{equation}
\frac{\delta}{\delta A_{a}^{i}(x)}\ U(\gamma,A)
= \int_{\gamma} ds\ \frac{dx^{a}(s)}{ds}\	\delta^{3}(\gamma(s),x)
\ U(\gamma(0,s),A)\ \tau^{i}\ U(\gamma(s,1),A).
\label{derivative}
\end{equation}
Here, $s\in[0,1]$ is a coordinate along the curve $\gamma: s \to 
x^{a}(s)$ and the curves $\gamma(0,s)$ and $\gamma(s,1)$ are the 
two segments in which the point with coordinate $s$ cuts 
$\gamma$.  It follows that the derivative of a matrix in the 
representation $j$ is
\begin{equation}
\frac{\delta}{\delta A_{a}^{i}(x)}\ j[U(\gamma,A)]
= \int_{\gamma} ds\ \frac{dx^{a}(s)}{ds}\	\delta^{3}(\gamma(s),x)
\ \ j[U(\gamma(0,s),A)]\ \tau^{i}_{(j)}\ j[U(\gamma(s,1),A)],
\label{jderivative}
\end{equation}
where $\tau^{i}_{(j)}$ are the generators of the spin $j$ 
representation of $SU(2)$.  Let us isolate $j[U(\gamma,A)]$ 
($\gamma$ being the link that crosses $\Sigma$) in the state and 
write the state $\Psi_{s}(A)$ as 
\begin{equation}
	\Psi_{s}(A) = \Psi_{(s-\gamma)}^{lm}(A)\ j[U(\gamma,A)]_{lm},
\end{equation}
where $l$ and $m$ are indices in the Hilbert space of the 
representation associated to $\gamma$.  From (\ref{E}), the 
action of the operator $E^{i}(\Sigma)$ on $\Psi_{s}$ is
\begin{equation}
	E^{i}(\Sigma)\ \Psi_{s}(A) = -i\int_{\Sigma}d\sigma^{1}d\sigma^{2} 
	\epsilon_{abc} 
	\frac{\partial x^{a}}{\partial \sigma^{1}} 
	\frac{\partial x^{b}}{\partial \sigma^{2}} \ 
	 \Psi_{(s-\gamma)}^{lm}(A)\ 
	  \frac{\delta}{\delta A_{c}^{i}(x)}
	   j[U(\gamma,A)]_{lm}. 
\end{equation}
Using (\ref{jderivative}), we obtain 
\begin{eqnarray}
	E^{i}(\Sigma) \Psi_{s}(A) &=& -i
	\int_{\Sigma}d\sigma^{1}d\sigma^{2} 
  \int_{\gamma} ds \ 
  	\epsilon_{abc} 
	\frac{\partial x^{a}(\vec\sigma)}{\partial \sigma^{1}} 
	\frac{\partial x^{b}(\vec\sigma)}{\partial \sigma^{2}} \ 
	\frac{dx^{c}(s)}{ds}\ 
	\delta^{3}(\gamma(s),x(\vec\sigma))
	\nonumber \\ && \times \ \ 
	 \Psi_{(s-\gamma)}^{lm}(A) \ 
\left(j[U(\gamma(0,s),A)]\ \tau^{i}_{(j)}\ 
j[U(\gamma(s,1),A)]\right)_{lm}. 
\label{full}
\end{eqnarray}
Remarkably, the three partial derivatives combine to produce the 
jacobian for the change of integration coordinates from 
$(\sigma^{1},\sigma^{2},s)$ to $x^{1},x^{2},x^{3}$.  If this 
jacobian is non vanishing, we perform the change of integration 
coordinates and then we can integrate away the delta function, 
obtaining
\begin{equation}
	E^{i}(\Sigma)\ \Psi_{s}(A) = -i 
	 \Psi_{(s-\gamma)}^{lm}(A) \  
\left(j(U[\gamma(0,s),A)]\ \tau^{i}_{(j)}\ 
j[U(\gamma(s,1),A)]\right)_{lm}. 
\end{equation} 
Thus, the effect of the operator $E^{i}(\Sigma)$ on the state 
$\Psi_{s}(A)$ is simply the insertion of the 
matrix $-i\tau^{i}_{(j)}$ in the point corresponding to the 
puncture. 

If, on the other hand, the jacobian vanishes, then the entire 
integral vanishes.  This happens if the tangent to the loop 
$\frac{dx^{a}(s)}{ds}$ is tangent to the surface.  In particular, 
for instance, this happens if the loop $\gamma$ lies entirely on 
the surface, in which case the puncture is not an isolated point.  
Therefore only isolated punctures contribute to $E^{i}(\Sigma) 
\Psi_{s}(A)$.

The key of the above computation is the analytical expression for 
the (integer) intersection number $I[\Sigma, \gamma]$ between a 
surface $\Sigma$ and a loop $\gamma$
\begin{equation}
	I[\Sigma, \gamma] = 
	\int_{\Sigma}d\sigma^{1}d\sigma^{2} 
    \int_{\gamma} ds \ 
  	\epsilon_{abc} 
	\frac{\partial x^{a}(\vec\sigma)}{\partial \sigma^{1}} 
	\frac{\partial x^{b}(\vec\sigma)}{\partial \sigma^{2}} \ 
	\frac{dx^{c}(s)}{ds}\ 
	\delta^{3}(\gamma(s),x(\vec\sigma)). 
\end{equation}
This integral is independent of the coordinates $\vec\sigma, s$ 
and $x^{a}$, and yields an integer: the (oriented) number of 
punctures. (The sign is determined by the relative orientation 
of surface and loop.)

Finally, it is easy to see that what happens if the surface 
$\Sigma$ and the spin network $s$ intersect in more than one 
puncture (along the same or different links).  In this case 
$E^{i}(\Sigma) \Psi_{s}(A)$ is a sum of one term per puncture -- 
each term being given by the insertion of a $\tau$ matrix.  
Finally, a bit more care is required for the computation of 
$E^{i}(\Sigma) \Psi_{s}(A)$ when the punctures are also nodes of 
$s$.  This case is discussed in Appendix B. 

\subsection{The operator $A(\Sigma)$}

Clearly $E^{i}(\Sigma)$ is not an $SU(2)$ gauge invariant 
operator.  In fact, we have seen in the previous section that 
acting on spin network states, which are in ${\cal H}_{0}$, it 
gives states which, in general, are not in ${\cal H}_{0}$.  
Consider the operator
\begin{equation}
	E(\Sigma)=\sqrt{E^{i}(\Sigma)E^{i}(\Sigma)}. 
	\label{Edef}
	\end{equation}
Acting on a state $\Psi_{s}$ that intersects $\Sigma$ only once, it 
gives
\begin{eqnarray}
	E^{2}(\Sigma)\Psi_{s}(A) &=& - 
	 \Psi_{(s-\gamma)}^{lm}(A) \ 
\left(j(U[\gamma(0,s),A))\  \tau^{i}_{(j)}\ \tau^{i}_{(j)} \ 
j[U(\gamma(s,1),A)]\right)_{lm}
\nonumber \\
	&=&
	 \Psi_{(s-\gamma)}^{lm}(A)
\left(j[U(\gamma(0,s),A))\ j(j+1)\ j[U(\gamma(s,1),A)]\right)_{lm} 
\nonumber \\
	&=&
	j(j+1)\ \Psi_{s}(A). 
\end{eqnarray}
(The relative orientation of the surface and the link becomes  
irrelevant because of the square.) Thus
\begin{equation}
		E(\Sigma)\ \Psi_{s}(A) 
	= \sqrt{j(j+1)}\  \Psi_{s}(A). 
	\label{jjpu}
\end{equation}
This is a nice result: not only $E(\Sigma)\Psi_{s}$ is in ${\cal 
H}_{0}$, the space of the $SU(2)$ gauge invariant states, but we 
even have that the spin network state $\Psi_{s}$ is an 
eigenstate.  Both properties, however, hold only because there 
is a {\em single\/} puncture.  It is easy to see that if there 
are two (or more) punctures, the cross terms spoil the gauge 
invariance of $E^{2}(\Sigma)\Psi_{s}$.  We are now going to 
define a $SU(2)$ invariant operator by eliminating the cross 
terms.  To this purpose, let $\Sigma^{(n)}=(\Sigma^{(n)}_{k}, 
k=1\ldots n)$ be a sequence of increasingly fine partitions of 
$\Sigma$ in $n$ small surfaces.  We define
\begin{equation}
A(\Sigma) = \lim_{n\to\infty} \sum_{k} E(\Sigma^{(n)}_{k}).
\label{Adef}
\end{equation}
For every given spin network state $\Psi_{s}$, there is a 
partition $(n)$ sufficiently fine for which each puncture $i$ 
falls on a different small surface 
$\Sigma_{k}$.  For these $k$, equation (\ref{jjpu}) holds, while 
the other $E^{i}(\Sigma_{k})$ vanish.  From that $(n)$ on, the 
limit is trivial.  And therefore 
\begin{equation}
	A(\Sigma)\ \Psi_{s}=\sum_{i\in(s\cap\Sigma)} \sqrt{j_{i}(j_{i}+1)}
	\  \Psi_{s},
	\label{A}
\end{equation}
where the index $i$ labels the punctures.  This formula holds for 
all cases except for the case in which a node of $p$ is on 
$\Sigma$, for which see Appendix B.

The operator $A(\Sigma)$ is well defined on ${\cal H}_{0}$. It 
is diagonalized by spin network states. It is self-adjoint.  Its 
spectrum includes the main sequence given by (\ref{A}), 
namely 
\begin{equation}
	A = \sum_{i} \sqrt{j_{i}(j_{i}+1)}.
	\label{spettrom}
\end{equation}
The full spectrum is computed taking also the cases with 
nodes into account (see Appendix B).  It is given by
\begin{equation}
	A = \sum_{j_{i}^{u},j_{i}^{d},j_{i}^{t}} 
	\sqrt{j^{u}_{i}(j^{u}_{i}+1) + j^{d}_{i}(j^{d}_{i}+1)
	-1/2 j^{t}_{i}(j^{t}_{i}+1)}. 
\label{sptutto}
\end{equation}

In conclusion, for each surface $\Sigma$ in $M$ there is a 
self-adjoint operator $A(\Sigma)$, with spectrum (\ref{sptutto}), 
which is defined by equations (\ref{E}), (\ref{Edef}) and 
(\ref{Adef}), and which is diagonalized by a basis of spin 
network states.

\section{Physics} \label{Phys}

We now provide a physical interpretation to the construction of 
the previous section.  General relativity can be seen as a 
diffeomorphism invariant theory for a $SU(2)$ connection 
$A_{a}^{i}(x)$ \cite{Barbero,Ashtekar}.  
In this formalism, the conjugate momentum 
observables $E^{a}_{i}(x)$ must be interpreted as the inverse 
densitized triad.  That is, the three-dimensional metric 
$g_{ab}(x)$ is defined by
\begin{equation}
	(\det g) g^{ab}(x) = E^{a}_{i}(x) E^{b}_{i}(x). 
\label{g}
\end{equation}
The space $\cal A$ defined in Section \ref{Math} has thus a 
natural interpretation as the configuration space of general 
relativity in this formalism.  The two (kinematical)  gauge groups 
of the theory, namely local $SU(2)$ and the group of the 
diffeomorphisms act naturally in $\cal A$.

To quantize the theory canonically, we need a Hilbert space of 
states on which an algebra of observables is represented.  The 
simplest possibility is to consider a space of functions over 
configuration space.  The Hilbert space $\cal H$ is an example of 
such a space.  It represents an interesting candidate 
particularly because it carries natural unitary representations of the 
two invariance groups of the theory.

The analogous choice would not be viable in a conventional 
Yang-Mills theory, because $\cal H$ is a ``far too big'' state 
space (in fact, non-separable), obtained by assigning finite norm 
to far ``too singular'' functions, having domain of dependence on 
one-dimensional curves.  But the presence of the diffeomorphism 
group changes the game drastically, because the physical states 
invariant under (a suitable quantum version of) the 
diffeomorphisms are {\em de facto\/} ``smeared all over the 
manifold'', and form a separable Hilbert space.  On the other 
hand, the state spaces on which conventional quantum field theory 
is built (say, Fock space) do not carry a sensible unitary 
representation of the diffeomorphism group.  Their structure 
relies heavily on the existence of a background metric, and the 
difficulty of making sense of background independent notions in 
field theory (or string theory) is well known (see for instance 
the discussion in \cite{Witten}).  The physical hypothesis that 
loop quantum gravity explores is that the correct mathematics for 
describing in a non-perturbative and background independent 
fashion a diffeomorphism invariant theory --a theory {\em of\/} 
spacetime, and not a theory {\em in\/} spacetime-- must be 
searched in rather different realms than the usual splendor of 
{\em conventional\/} quantum field theory.

Thus we interpret the states $\Psi(A)$ in $\cal H$ as quantum 
states of the gravitational field (prior to imposing the quantum 
constraints).  The conjugate momentum operator 
$\frac{\delta}{\delta A_{a}^{i}(x)}$ is then to be interpreted as 
the quantum operator corresponding to the inverse densitized 
triad $E^{a}_{i}(x)$ (the two have correctly, of course, the same 
geometrical transformation properties).  More precisely, the 
canonical commutation relations between $A$ and $E$ are
\begin{equation}
 	\{A_{a}^{i}(x), E^{b}_{j}(y) \} = G\ \delta_{a}^{b}\ 
 	\delta_{j}^{i}
\ \delta^{3}(x,y). 
\end{equation}
(The presence of the Newton constant $G$ is due to the fact that 
the momentum conjugate to $A$, which is defined as the derivative 
of the action with respect to $\frac{dA}{dt}$, is actually 
$\frac{1}{G}E$.  The Newton constant enters the canonical 
formalism and the quantum theory precisely in the same manner the 
mass of a free particle does.)  Therefore the operator 
corresponding to $E$ is $-i \hbar G \frac{\delta}{\delta 
A}$.  This operator --suitably smeared-- together with the 
holonomy of $A$, forms a operator algebra which quantizes the 
corresponding classical Poisson algebra.

What is then the physical interpretation of the operator 
$A(\Sigma)$ constructed above?  The classical limit of 
$A(\Sigma)$ is obtained replacing $-i \hbar G 
\frac{\delta}{\delta A}$ with $E$ in its definition.  (Let us put 
$\hbar G = 1$ for a moment.)  We obtain from 
(\ref{Adef}), (\ref{Edef}) and (\ref{E})
\begin{eqnarray}
A(\Sigma) &=& \lim_{n\to\infty} \sum_{k} 
\sqrt{E^{i}(\Sigma^{(n)}_{k})E^{i}(\Sigma^{(n)}_{k})}. 
\label{uno} \\ 
 E^{i}(\Sigma^{(n)}_{k})  &=& 
 \int_{\Sigma^{(n)}_{k}}d\sigma^{1}d\sigma^{2} \ 
	\epsilon_{abc} \frac{\partial x^{a}}{\partial \sigma^{1}} 
	\frac{\partial x^{b}}{\partial \sigma^{2}} \ 
	E^{ci}(x(\vec\sigma)). 
	\label{inte}
\end{eqnarray}
For a sufficiently fine triangulation $(n)$, $\Sigma^{(n)}_{k}$ is 
arbitrarily small.  We can thus replace the integration in 
(\ref{inte}) with the value of the integrand in a point 
$x_{k}$ in $\Sigma^{(n)}_{k}$, times the coordinate area   
$a(\Sigma^{(n)}_{k})$ of the small surface 
\begin{equation}
E^{i}(\Sigma^{(n)}_{k})  =  a(\Sigma^{(n)}_{k}) \ 
\epsilon_{abc} \frac{\partial x^{a}}{\partial \sigma^{1}} 
\frac{\partial x^{b}}{\partial \sigma^{2}} 
E^{ci}(x_{k}).
\end{equation}
Inserting this in (\ref{uno}), we obtain precisely the 
definition of Riemann integral, yielding 
\begin{equation}
A(\Sigma) = \int_{\Sigma} d\sigma^{1} d\sigma^{2} 
\sqrt{\epsilon_{abc} \frac{\partial x^{a}}{\partial \sigma^{1}} 
\frac{\partial x^{b}}{\partial \sigma^{2}} 
E^{ci}(x(\vec\sigma))\ 
\epsilon_{def} \frac{\partial x^{d}}{\partial \sigma^{1}} 
\frac{\partial x^{e}}{\partial \sigma^{2}} 
E^{fi}(x(\vec\sigma))}. 
\end{equation}
This expression is independent from the choice of both $x$ 
and $\sigma$ coordinates. The easiest way to evaluate it is 
to choose $x^{3}=0$ on $\Sigma$ and $\sigma^{1}=x^{1}, 
\sigma^{2}=x^{2}$. We obtain 
\begin{equation}
A(\Sigma) 
= \int_{\Sigma} d\sigma^{1} d\sigma^{2} 
\sqrt{E^{3i}(x) E^{3i}(x)} 
= \int_{\Sigma} d\sigma^{1} d\sigma^{2} 
\sqrt{(\det g) g^{33}}. 
\end{equation}
where we have used (\ref{g}). Using the formula for the 
inverse of a matrix, we have 
\begin{equation}
A(\Sigma) 
= \int_{\Sigma} d\sigma^{1} d\sigma^{2} 
\sqrt{g_{11}g_{22}-g_{12}g_{12}} 
= \int_{\Sigma} d\sigma^{1} d\sigma^{2} 
\sqrt{\det {}^{2}g}. 
\end{equation}
where ${}^{2}g$ is the 2d metric induced by $g_{ab}$ on 
$\Sigma$. The last equation shows that the result is 
covariant and is immediately recognized as the physical 
area of $\Sigma$.  In conclusion, and restoring physical 
units, we have 
\begin{equation}
	\hbar G\ A(\Sigma) = {\rm Area\ of}\ \Sigma . 
\end{equation}
The area depends on the metric and the metric is the 
gravitational field.  In quantum gravity, the gravitational field 
is given by a physical operator, and therefore the area is 
represented by an operator as well.  According to standard 
quantum mechanics rules, the spectrum of the operator corresponds 
to the possible outcomes of individual measurements of the 
corresponding physical quantity.  Therefore the hypothesis that 
the quantum theory of gravity can be based on the Hilbert 
space $\cal H$ --which is the basic hypothesis of loop quantum 
gravity-- yields the physical result that Planck scale 
measurements of the area can yield only quantized outcomes.  
These are labeled (see (\ref{fullspectrum})) by $n$-tuplets of 
triplets of half integers $(j_{i}^{u}, j_{i}^{d}, j_{i}^{t}), 
i=1,\ldots,n$ and given by
\begin{equation}
A = \frac{1}{2}\ \hbar G\ \sum_{i} 
\sqrt{2 j_{i}^{u}(j_{i}^{u}+1) + 2 j_{i}^{d}(j_{i}^{d}+1) - 
j_{i}^{t}(j_{i}^{t}+1)} .
\label{spettro}
\end{equation}
For $j_{i}^{t}=0$ and $j_{i}^{u}=j_{i}^{d}=j_{i}$, this reduces to 
the spectrum
\begin{equation}
A = \hbar G\ \sum_{i}\ \sqrt{j_{i}(j_{i}+1)}.
\label{spettrocorto}
\end{equation}

A few comments are needed, in order to qualify this result. 
\begin{itemize}

\item The overall multiplicative factor in front of the right hand 
side of (\ref{spettro}) is uncertain.  In fact, it has been 
noted that the loop quantization of a $SU(2)$ theory contains an 
arbitrary parameter, the Immirzi parameter, which rescales the 
spectrum.  See \cite{Immirzi} for details.  

\item The operator $A(\Sigma)$ is invariant under $SU(2)$ gauge 
transformations, but {\em not\/} under three or four dimensional 
diffeomorphisms.  Therefore, strictly speaking it is not an 
observable of the theory, and we cannot directly give its 
spectrum physical meaning.  The failure of $A(\Sigma)$ to be 
diff-invariant is a consequence of the fact that the area of an 
abstract surface defined in terms of coordinates is not a diff 
invariant concept.  In fact, physical measurable areas in general 
relativity correspond to surfaces defined by physical degrees of 
freedom, for instance matter (the area of the surface this table) 
or the gravitational field itself (the area of an event horizon).  
However, it is reasonable to expect that the fully gauge 
invariant operator corresponding to a physically defined area 
(say defined by matter) has precisely the same mathematical form 
as the non gauge invariant operator studied here.  The reason is 
that one can use matter degrees of freedom to gauge-fix the 
diffeomorphisms -- so that a non-diff-invariant quantity in pure 
gravity corresponds to a diff-invariant quantity in a 
gravity+matter theory.  This expectation has been confirmed 
explicitly in a number of cases \cite{Area,Leearea}.

\item Objections have been raised about the last point.  See in 
particular \cite{Amelino}.  Some objections are based on the 
intuition that the position of the matter defining the surface 
could be subjected to quantum fluctuations, preventing the 
possibility of defining a sharp surface.  This objection is incorrect.  
Neither the position of matter, nor the area of a surface, have 
physical independent reality.  It is only the gravitational field 
in the location determined by the matter, or, the other way 
around, the location of the matter in the gravitational field, 
that have physical reality.  The two do not form independent sets 
of degrees of freedom subjected to independent quantum 
fluctuations.  See \cite{observable} for a detailed discussion of 
this point.

\item A result analogous to the one for the area holds for 
the volume. 

\item There are many spin network bases, because a basis in 
each $H_{p}$ must be selected in order to have a fully determined 
choice.  For fixed $\Sigma$, we can choose a basis (in fact, many 
bases) that diagonalizes $A(\Sigma)$.  However, we cannot choose a 
basis that diagonalizes $A(\Sigma)$ for every $\Sigma$.  To see 
how this may happen, consider a the 4-valent node  
described in Appendix A. The area of a surface $\Sigma_{1}$ that 
separates $\gamma_{1}$ and $\gamma_{2}$ from $\gamma_{3}$ and 
$\gamma_{4}$ is diagonalized by the basis (\ref{base1}), while 
the area of a surface $\Sigma_{2}$ that separates $\gamma_{1}$ 
and $\gamma_{3}$ from $\gamma_{2}$ and $\gamma_{4}$ is 
diagonalized by the basis (\ref{base2}).  Thus, $A(\Sigma_{1})$ 
and $A(\Sigma_{2})$ do not commute, and there is no basis that 
diagonalizes all area operators. 

\item The spectrum of the area operator is composed by two 
sequences.  A main sequence (\ref{spettrocorto}), and a 
secondary sequence, given by (\ref{spettro}) with $j^{(t)}\ne 
0$, generated by the states with nodes on the surface.  Based on 
the intuition that such states are --in a sense-- degenerate, 
doubts have been raised on whether the secondary sequence must 
be considered a prediction of the theory on the same ground as 
the main sequence (\ref{spettrocorto}).

\item Obviously, direct experimental verification of the 
predicted spectrum (\ref{spettro}) is far outside the present 
possibilities.  Possibilities of empirical corroboration, if any, 
are likely to be indirect.  For instance, future observations of 
the cosmic background gravitational radiation might perhaps 
reveal an imprint of the spectrum.  This possibility has not yet 
been investigated.  On the other hand, the result (\ref{spettrocorto}) 
is the basis of all applications of loop quantum gravity to black 
hole thermodynamics \cite{Helvetica,bh}.

\end{itemize}

In summary, we have defined the unconstrained state space of 
quantum gravity in the loop representation, and the basic 
operators of the theory.  We have then defined an operator 
$A(\Sigma)$, associated to each surface $\Sigma$, and computed 
its spectrum.  We have finally shown that the classical limit of 
$A(\Sigma)$ is the geometrical area of the surface $\Sigma$.  
This leads to the physical hypothesis that the spectrum of 
$A(\Sigma)$ describes the quantizations of the physically 
measurable area.  This constructions puts the old suggestion that 
at the Planck scale geometry is discrete (in the quantum sense) 
in a precise framework and gives it a precise and non trivial 
quantitative form.   

For a discussion of the rest of the loop quantum gravity and 
the problems still open in the theory, and for detailed 
references, see \cite{Loops}. 

\section*{Acknowledgments}

We thank Don Marolf for a useful exchange.  This work has been 
supported by NSF Grant PHY-95-15506.

\section*{Appendix A: Details on spin network states}

We add a few details in order to clarify the construction of 
the previous subsection.  

\subsection{$SU(2)$}

We begin by recalling some basic facts about $SU(2)$ 
representation theory, in order to fix our notation.  Let 
$U^{A}{}_{B}$, where $A,B=0,1$ be a $SU(2)$ matrix.  These 
matrices act on $C^{2}$ vectors $v^{A}$, and leave the unit 
antisymmetric objects $\epsilon^{AB}$ and $\epsilon_{AB}$ (where 
$\epsilon^{01}=\epsilon_{01}=1$) invariant.  Using 
$\epsilon_{AB}$ (from, say, the left), we can lower the first 
index of the matrix $U$, defining $U_{AB}=\epsilon_{AC} 
U^{C}{}_{B}$.  The irreducible representations of $SU(2)$ are 
labeled by their spin $j$, which is a nonnegative half-integer.  
The spin-$j$ representation has dimension $2j+1$ and can be 
obtained as the symmetrized tensor product of $2j$ copies of the 
fundamental (2-dimensional) representation.  Thus the Hilbert 
space $H_{(j)}$ of the $j$ representation is the symmetrized 
tensor product of $2j$ copies of $C^{2}$.  We write vectors in 
$H_{(j)}$ (in ``spinor'' notation) as completely symmetric 
tensors with $2j$ indices $u^{A_{1}\ldots A_{2j}}$ (there are 
$2j+1$ unordered combinations of $2j$ zero's and one's).  For 
every $U\in SU(2)$, $j[U]$ is then obtained by symmetrizing the 
indices of the (tensor) product of $2j$ times $U^{A}{}_{B}$.
\begin{equation}
	j[U]^{A_{1}\ldots A_{2j}}{}_{B_{1}\ldots B_{2j}}
	=U^{(A_{1}}{}_{(B_{1}}\ldots U^{A_{2j})}{}_{B_{2j})}. 
\end{equation}
Again, we can lower all upper indices of $j[U]$ with 
$\epsilon_{AB}$'s. 

\subsection{The invariant tensors $v$}

The invariant tensors $v$ defined in Section \ref{Math} are 
$SU(2)$ invariant vectors in the tensor product 
$H_{(j_{1})}\otimes \ldots \otimes H_{(j_{n})}$ of the 
representations $j_{1},\ldots,j_{n}$ associated to the links 
$\gamma_{1},\ldots,\gamma_{n}$ that meet at a $n$-valent node 
$p$.  Thus they have the form $v^{(A_{1}\ldots 
A_{2j_{1}}),\ldots,(D_{1}\ldots D_{2j_{n}})}$.  Since the only 
invariant object is $\epsilon^{AB}$, they are given by suitably 
symmetrized sequences of $\epsilon^{AB}$'s.  Recall we denote the 
space of such vectors (the spin zero subspace of the tensor 
product) as $H_{p}$.

$H_{p}$ has dimension zero if the valence of $p$ is one.  If $p$ 
is bi-valent, $H_{p}$ has dimension zero unless the two links are 
associated to the same representation.  In this case, the only 
invariant tensor is
\begin{equation}
	v^{A_{1}\ldots A_{2j},\ B_{1}\ldots B_{2j}}
	=\epsilon^{(A_{1}(B_{1}}\ldots\ \epsilon^{A_{2j})B_{2j})}. 
\label{due}
\end{equation}
The most interesting case is if $p$ is trivalent, then $H_{p}$ 
has dimension zero unless the three representations satisfy the 
Clebsh-Gordon condition $|j_{2}-j_{3}|\le j_{1}\le j_{2}+j_{3}$ 
(the angular momentum addition condition: each one of three 
representations is in the tensor product of the other two).  If 
they do satisfy these conditions, then $v$ is the Wigner 3-$j$ 
symbol, namely the symmetric form of the Clebsh-Gordon 
coefficients, which can be written as
\begin{equation} 
v^{A_{1}\ldots A_{2j_{1}},\ B_{1}\ldots B_{2j_{2}},\ C_{1}\ldots 
C_{2j_{3}}}	
= \epsilon^{A_{1}B_{1}} \ldots 
\epsilon^{A_{a}B_{a}} \ \ 
\epsilon^{B_{a+1}C_{1}} \ldots 
\epsilon^{B_{a+b}C_{b}}\ \ 
\epsilon^{A_{a+1}C_{b}} \ldots 
\epsilon^{A_{a+c}C_{b+c}}
\label{tre}
\end{equation}
where complete symmetrization in the $A$ indices, in the $B$ indices 
and in the $C$ indices is understood, and
\begin{equation}
	2j_{1}=a+c, \ \ \ 
	2j_{2}=a+b, \ \ \ 
	2j_{3}=b+c. 
	\label{cg}
\end{equation}
Notice that the existence of three integers $a$, $b$ and $c$ 
satisfying (\ref{cg}) is equivalent to the Clebsh-Gordon 
condition on the three spins $j_{1}, j_{2}, j_{3}$.  Indeed, the 
Clebsh-Gordon condition is equivalent to the possibility of 
rooting lines across a fork: if we have a triple node  
with $2j_{1}, 2j_{2}, 2j_{3}$ lines coming along each of the 
three links $\gamma_{1}, \gamma_{2}, \gamma_{3}$, then 
$j_{1}, j_{2}, j_{3}$ satisfy the Clebsh-Gordon condition if and 
only if the lines can be consistently rooted.  Then $a$ lines 
connect  $\gamma_{1}$ and $\gamma_{2}$, $b$ lines connect $\gamma_{2}$ and 
$\gamma_{3}$ and $c$ lines connect $\gamma_{3}$ and $\gamma_{1}$.

For higher valence $n$, the $n$-indices tensor $v$ can be 
constructed by contracting 3-$j$ symbols.  It is easier at this 
point to shift to vector notation.  That is, to write 
$u^{A_{1},\ldots A_{2j}}$ as $u_{(j)}^{m}$ where 
$m=1\ldots(2j+1)$.  Then we write the representation matrices as 
$j[U]^{a}{}_{b}$ (and $j[U]_{ab}$) and we write (\ref{due}) and 
(\ref{tre}) as $q_{(j_{1})(j_{2})}^{lm}$ and $v^{lmn}_{(j_{1}) 
(j_{2}) (j_{3})}$. The $4$-index $v$ 
have the form $v^{lmnp}_{(j_{1}) (j_{2}) (j_{3})(j_{4})}$. 
A basis in the space of these objects is 
given as 
\begin{equation}
	v^{lmnp}_{(j_{1}) (j_{2}) (j_{3})(j_{4})}=
	c^{j}\ 
	e^{lmnp}_{j}{}_{(j_{1}) (j_{2}) (j_{3})(j_{4})}
\end{equation}
where the basis elements are
\begin{equation} 
e^{lmnp}_{j}{}_{(j_{1}) (j_{2}) (j_{3})(j_{4})}=
v^{lmr}_{(j_{1}) (j_{2}) (j)}\ \  q^{(j)}_{rs} \ \ 
v^{snp}_{(j)(j_{3}) (j_{4})}. 
\label{four}
\end{equation}
Here $j$ is any representation such that the Clebsh-Gordon 
condition are satisfied (between $j_{1}, j_{2}, j$ as well as 
between $j, j_{3}, j_{4}$).  Since there is a finite number of 
representations $j$ satisfying these Clebsh Gordon conditions, 
the $v$ span a finite dimensional space, which is $H_{p}$.  A 
direct calculation shows that the basis vectors (with different 
$j$'s) (\ref{four}) are orthogonal to each other.  We can view 
equation (\ref{four}) as representing the expansion of a 
4-valent node into two 3-valent nodes, joined by a 
``virtual'' link with color $j$.  Notice that in (\ref{four}) we 
could have grouped the 4 representations in couples differently.  
Namely, there exists another basis
\begin{equation} 
a^{lmnp}_{j}{}_{(j_{1}) (j_{2}) (j_{3})(j_{4})}=
v^{lnr}_{(j_{1}) (j_{3}) (j)}\ \  q^{(j)}_{rs} \ \ 
v^{smp}_{(j)(j_{2}) (j_{4})}.
\label{base1}
\end{equation}
According to the recoupling theorem, the matrix of the change 
of basis is given by the Wigner 6-$j$ symbols
\begin{equation} 
a^{abcd}_{j}{}_{(j_{1}) (j_{3}) (j_{2})(j_{4})}=
\sum_{k} \ \left(
\begin{array}{ccc}j_{1}&j_{2}& j\\ j_{3}&j_{4}& k\end{array}
\right)\ \ 
e^{abcd}_{k}{}_{(j_{1}) (j_{2}) (j_{3})(j_{4})}. 
\label{base2}
\end{equation}

The technique of expanding a node in 3-valent nodes 
joined by virtual links can be used for constructing the 
basis of $H_{p}$ in general. For every $n$-valent node, we write a 
``virtual'' tree-like (no closed loops) graph with $n$ open ends. 
The dimension of  $H_{p}$ is given by counting the number of 
colorings of the virtual links which are (Clebsh-Gordon-) 
compatible at all virtual 3-valent nodes.  Each topologically 
distinct virtual decomposition defines a basis in $H_{p}$. 

\subsection{Holonomies}

We recall that given a loop $\gamma:s\in[0,1]\to \gamma(s)\in M$ 
and a connection $A$ on $M$, the parallel propagator 
$U(\gamma,A)$ is the value for $s=1$ of the $SU(2)$ matrix $U(s)$ 
that solves the differential equation
\begin{equation}
\frac{dU(s)}{ds} + \frac{d\gamma^a}{ds} A_a(\gamma(s)) U(s) = 0
\label{holonomy}
\end{equation}
with the initial condition $U(0)=1$, the identity element of 
$SU(2)$. The solution of this equation is formally written as 
\begin{equation}
U(\gamma,A) = {\cal P}\, 
e^{\int^1_{0}\,A_a(\gamma(s))\ \frac{d\gamma^a}{ds}\ ds}, 
\end{equation}
and can be given explicitly as a power series in $A$ by expanding 
the exponential in powers and path-ordering the powers of $A$ 
along $\gamma$.  (That is ${\cal P}\, A(\gamma(s'))A(\gamma(s)) 
\equiv A(\gamma(s)) A(\gamma(s'))$ if $s<s'$.)

\subsection{Spin network states}

Given a spin network $s=(\Gamma,j_{i},v_{r})$, the (representative 
in the representation $j_{i}$ of the) parallel propagator of the 
connection $A$ along each link $\gamma_{i}$ defines the matrix 
$j[U(\gamma_{i},A)]_{lm}$.   By contracting these matrices with 
the invariant tensors $v_{r}$ at each node $p_{r}$, we obtain the spin 
network state $\Psi_{s}(A)$.   

As a simple example, consider a graph $\Gamma$ formed by two 
nodes $p$ and $q$ in $M$ joined by three links $\gamma_{1}, 
\gamma_{2}, \gamma_{3}$.  Then
\begin{equation}
	\Psi_{(\Gamma,j_{1}j_{2}j_{3})}=
	v^{lmn}_{(j_{1}) (j_{2}) (j_{3})}\ 
	j[U(\gamma_{1},A)]_{lp}\ 
	j[U(\gamma_{2},A)]_{mq}\ 
	j[U(\gamma_{3},A)]_{nr}\ 
	v^{pqr}_{(j_{1}) (j_{2}) (j_{3})}. 
\label{example}
\end{equation}
(In this example, there is no need of indicating explicitly the 
coloring of the nodes, because they are both trivalent and 
therefore admit a unique coloring.)  Notice that in 
(\ref{example}) the pattern of the indices reflects the topology 
of the graph $\Gamma$.  In fact, spin network originated from 
Penrose's graphical notation for tensors.  For more details, see 
\cite{RovDP}. 

\section*{Appendix B: Nodes on $\Sigma$}

In this Appendix we discuss the action of $E^{i}(\Sigma)$ and 
$A(\Sigma)$ on $\Psi_{s}$ in the case that we have neglected in 
the main text, namely the case in which $s$ includes nodes which 
lies on $\Sigma$.  Let us assume that there is a single node $p$ 
of $s$ on $\Sigma$; the case in which there is more than one is 
then trivial.  Assume that $p$ is an $n-$valent node.  We must 
distinguish the links that meet in $p$ in three classes, which we 
denote ``up'', ``down'' and ``tangential''.  The tangential links 
are the ones that overlap with $\Sigma$ for a finite interval.  
The others links are naturally separated into two classes 
according to the side of $\Sigma$ they lie (which we arbitrarily 
label ``up'' and ``down'').  We chose a basis in $H_{p}$ as 
follows.  We decompose the $n$ valent node $p$ into a tree-like 
virtual graph (see Appendix A) by combining all ``up'' links into 
a virtual link ``up'', all ``down'' links into a virtual link 
``down'', and all ``tangential'' links into a virtual link 
``tangential''.  We call $j^{u}, j^{d}$ and $j^{t}$, 
respectively, the colors of these three virtual links.  These 
three virtual links then meet together, at the node 
$v_{(j^{u})(j^{d})(j^{t})}^{lmn}$.

By inspecting equation (\ref{full}), we see that the 
operator $E_{i}(\Sigma)$ fails to ``see'' the tangential 
links, because for these links the three derivatives in
\begin{equation}
	 	\epsilon_{abc} 
	\frac{\partial x^{a}(\vec\sigma)}{\partial \sigma^{1}} 
	\frac{\partial x^{b}(\vec\sigma)}{\partial \sigma^{2}} \ 
	\frac{dx^{a}(s)}{ds}
	\label{jac}
\end{equation}
are coplanar, and the above expression vanishes. Therefore 
the tangential links do not contribute. On the other hand, 
the other two set of links contribute (by symmetry) in equal 
fashion. Thus $E^{i}(\Sigma)$ inserts a $\tau_{(j^{u})}$ 
matrix in the ``up'' link plus a $\tau_{(j^{d})}$ 
matrix in the ``down'' virtual link.  In other words, if we 
write 
$\Psi_{s}=\Psi_{s}{}_{lmn}\ v_{(j^{u})(j^{d})(j^{t})}^{lmn}$, 
we have
\begin{equation}
E^{i}(\Sigma)\ \Psi_{s}
= \Psi_{s}{}_{lmn}\ 
\frac{1}{2}\left(\tau^{i}_{(j^{u})}{}_{p}^{l}\ \delta {}_{q}^{m}
 - \delta_{p}^{l}\  \tau^{i}_{(j^{d})}{}_{q}^{m} \right) \ 
 v_{(j^{u})(j^{d})(j^{t})}^{pqn}. 
\label{action}
\end{equation} 
Notice the minus sign coming from the inverse orientation of the 
jacobian (\ref{jac}) in the two cases. For a more complete  
derivation of  (\ref{action}), see \cite{nodes}. Since 
$v$ is an invariant tensor, 
\begin{equation}
	\left(\tau^{i}_{(j^{u})}+\tau^{i}_{(j^{d})}+
	\tau^{i}_{(j^{t})}\right)\ 
	v_{(j^{u})(j^{d})(j^{t})}=0,
\end{equation}
from which 
\begin{equation}
	(\tau^{i}_{(j^{u})}+\tau^{i}_{(j^{d})})^{2}
= (\tau^{i}_{(j^{t})})^{2}
\end{equation}
when acting on $v_{(j^{u})(j^{d})(j^{t})}$. 
Using this, simple algebra yields 
\begin{equation}
	(\tau^{i}_{(j^{u})}-\tau^{i}_{(j^{d})})^{2}
= 2(\tau^{i}_{(j^{u})})^{2}+2(\tau^{i}_{(j^{u})})^{2}- 
(\tau^{i}_{(j^{u})})^{2}=
2 j^{u}(j^{u}+1) + 2 j^{d}(j^{d}+1) - j^{t}(j^{t}+1).  
\end{equation}
In conclusion
\begin{equation}
A(\Sigma)\ \Psi_{s} = \frac{1}{2} 
\sqrt{2 j^{u}(j^{u}+1) + 2 j^{d}(j^{d}+1) - j^{t}(j^{t}+1)}\ \Psi_{s}.
\end{equation}
And in general 
\begin{equation}
A(\Sigma)\ \Psi_{s} = \frac{1}{2} \sum_{i\in(s\cap\Sigma)} 
\sqrt{2 j_{i}^{u}(j_{i}^{u}+1) + 
2 j_{i}^{d}(j_{i}^{d}+1) - j_{i}^{t}(j_{i}^{t}+1)}\ \Psi_{s}.
\label{fullspectrum}
\end{equation}
which is valid for all spin network states.  It includes also the 
case of a link $\gamma$ crossing $\Sigma$, because we can always 
rewrite $\gamma$ as two links joined by a 2-valent node on 
$\Sigma$, with $j^{t}=0$ and $j^{u}=j^{d}$.  In this case, 
(\ref{fullspectrum}) reduces to (\ref{A}).

\end{document}